\documentclass[twocolumn,prl,amsmath,superscriptaddress,showpacs,showkeys]{revtex4}

\usepackage[latin9]{inputenc}
\usepackage{units}
\usepackage{amsmath}
\usepackage{amssymb}
\usepackage{graphicx}
\usepackage{hyperref} 
\usepackage{color}
\usepackage{subfigure}
\usepackage{gensymb}
\usepackage{epstopdf}
\usepackage{enumerate}
\usepackage{mathbbol}

\begin{document}

\title{Magnetoresistance Anisotropy in Amorphous Superconducting Thin Films: A Site-Bond Percolation Approach}

\author{Elkana Porat and Yigal Meir}
\affiliation{Department of Physics, Ben-Gurion University, Beer Sheva 84105, Israel}

\date{\today}
\begin{abstract}
Recent measurements of the magnetoresistance (MR) of amorphous superconducting thin films in tilted magnetic fields have displayed several surprising experimental details, in particular a strong dependence of the MR on field angle at low magnetic fields, which diminishes and then changes sign at large fields. Using a generalized site-bond percolation model, that takes into account both the orbital and Zeeman effects of the magnetic field, we show that the resulting MR curves reproduce the main experimental features. Such measurements, accompanied by the corresponding theory, may be crucial in pinpointing the correct theory of the superconductor-insulator transition and of the MR peak in thin disordered films.
\end{abstract}



\maketitle

 \label{Intro}
The superconductor - insulator transition (SIT) in thin superconducting (SC) films has been observed 25 years ago \cite{Haviland1989,Hebard1990}, yet its nature is still under debate \cite{Goldman2010}, due to the interplay between superconductivity and disorder. While weak disorder has little effect on the SC state \cite{Anderson1959}, strong disorder  may lead to suppression of the SC state due to fluctuations of the
SC order parameter \cite{Kapitulnik1985, LeMa1985}. Indeed,
a  SIT has  been observed  upon tuning of e.g., the film thickness \cite{Haviland1989},
 external magnetic field \cite{Hebard1990} or disorder \cite{Shahar1992}.
Over the years, several paradigms for the transition have been put forward,
which can be broadly grouped into a pure bosonic paradigm, so called the ``dirty boson'' model
\cite{FisherGrinstinGirvin1990, Fisher1990, Galitski2005}, and variants of the percolation model
\cite{Kowal1994,Ghosal1998,Shimshoni1998,Mason1999,Dubi2006,Dubi2007,Dubi2008}.
A further insight into the nature of the transition was made possible due to magneto-resistance (MR) measurements in the normal phase \cite{Gantmakher1996,Sambandamurthy2004},
where giant resistance peak has been observed beyond the SIT, followed by a dramatic drop as the field is further increased.
This dramatic observation has been explained by a phenomenological percolation approach \cite{Dubi2006}, which emphasized the competition between the SC and the fermionic degrees of freedom, due to the persistence of SC islands (SCIs) into the insulating phase. An alternative explanation, based on the boson-only picture was put forward in Ref.\cite{Galitski2005}, where the role of the fermionic degrees of freedom was played by vortices.

\begin{figure}
\includegraphics[bb=70bp 210bp 490bp 640bp, clip, width=1\columnwidth,keepaspectratio]{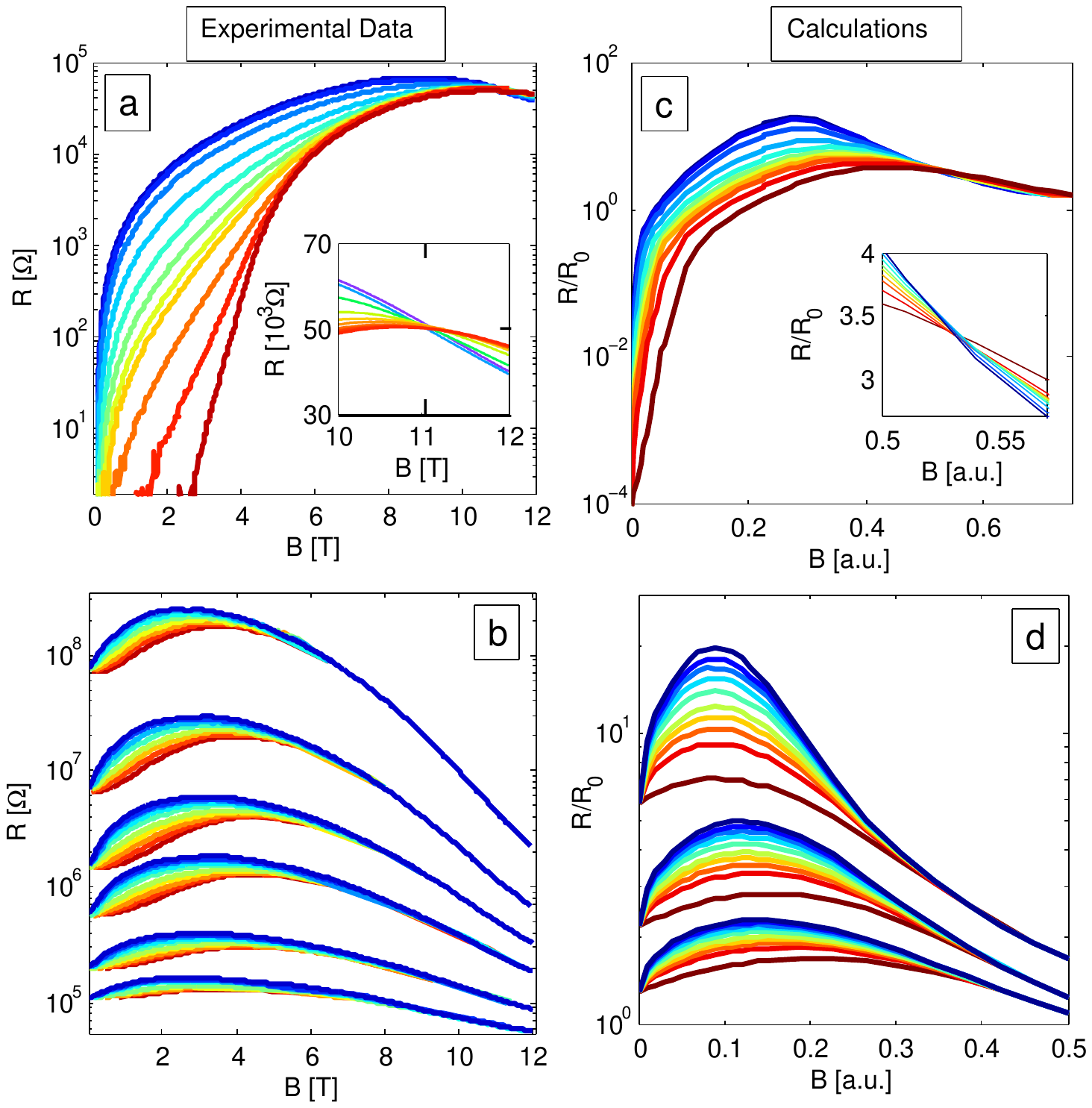}

\caption{\label{fig:experiments}
(a) MR isoterms from Ref. \onlinecite{Johansson2011}, of a sample which is a superconductor at zero field. The traces differ in angle $\theta$ of the magnetic field $B$, with respect to the plane of the sample, from $\theta = 0^\circ$, plotted in dark red, to $90^\circ$, plotted in dark blue.
Inset: The resistance is isotropic at the crossing point at $B = 11.02 T$.
(b) MR isoterms from Ref. \onlinecite{Shammass2012}, of a sample which is insulating at zero field.
Temperatures range between T =1 K and 0.3 K (bottom group to top group).
At each group of curves, colors represent different angles of $B$, from $\theta = 0^\circ$ (dark red) to $90^\circ$ (dark blue).
(c) Site-bond percolation results for low disorder (small $\Delta_{0}$, see text) - sample is initially in SC phase. The data were log-averaged over 100 realizations of a sample with 35X35 sites, at tilt angles similar to Fig. \ref{fig:experiments}(a),  with parameters:
$\Delta_{0}=0.25$, $W=0.6$, $E_{c}=8$, $T=1$, $x=0.5$, $\chi=0.05$, $n=0.25$.
(d) Site-bond percolation results for high disorder (larger $\Delta_{0}$) - sample is in the insulating phase.
The data were log-averaged over 100 realizations of a sample with 35X35 sites, at tilt angles similar to Fig. \ref{fig:experiments}(b), with model parameters:
$\Delta_{0}=0.5$, $W=0.6$, $E_{c}=8$, $x=0.5$, $\chi=0.005$, $n=0.25$ and temperatures $T=0.9, 2, 4$ (top to bottom).}
\end{figure}

While experiments indeed indicated the formation of SC puddles \cite{Kowal1994,Mason2001,Kowal2008,Sacepe2008,Kamlapure2013} and of critical (classical or quantum) percolation behavior \cite{Hebard1990,Yazdani1995,Markovic1998,Steiner2008,Schneider2012,Mehta2014}, it is clear that more experimental data are needed to establish the nature of the transition and of the insulating phase.
In recent years, detailed examinations of the MR dependence
on the direction of the field were performed \cite{Johansson2011,Shammass2012} (see Fig. \ref{fig:experiments}(a) and (b)).
The main observations are 
(1) highly anisotropic MR in the low field regime, reflected by high dependence of MR amplitude, SIT critical field $B_{c}$ and MR peak field $B_{max}$ on the field direction.
(2) Lower peak resistance was measured for shallower angles.
(3) The magnitude of the anisotropy decreases with the strength of the field, up to a point beyond the peak, where the MR curves seem to converge to isotropic MR, that is, angle independent resistance at a certain magnetic field intensity $B_{iso}$.
(4) In samples that are SC at zero field the anisotropy is reversed at higher fields,
i.e., higher resistance for shallower angles  \cite{Johansson2011}, while samples that are insulating at zero field depict nearly isotropic behavior for all $B>B_{iso}$ \cite{Shammass2012}.

These new results are yet to be accounted for in any of the theoretical
pictures for the SIT, and thus are a key observation to discern the correct theory. In this letter we demonstrate that a phenomenological model within the framework of the percolation model
can explain these observations (see Fig. \ref{fig:experiments}). This gives further credibility to the percolation description of the SIT, and shed light on the different effects of the magnetic field.

\begin{figure}[Ht]
\centering 	
\includegraphics[width=0.95\columnwidth]{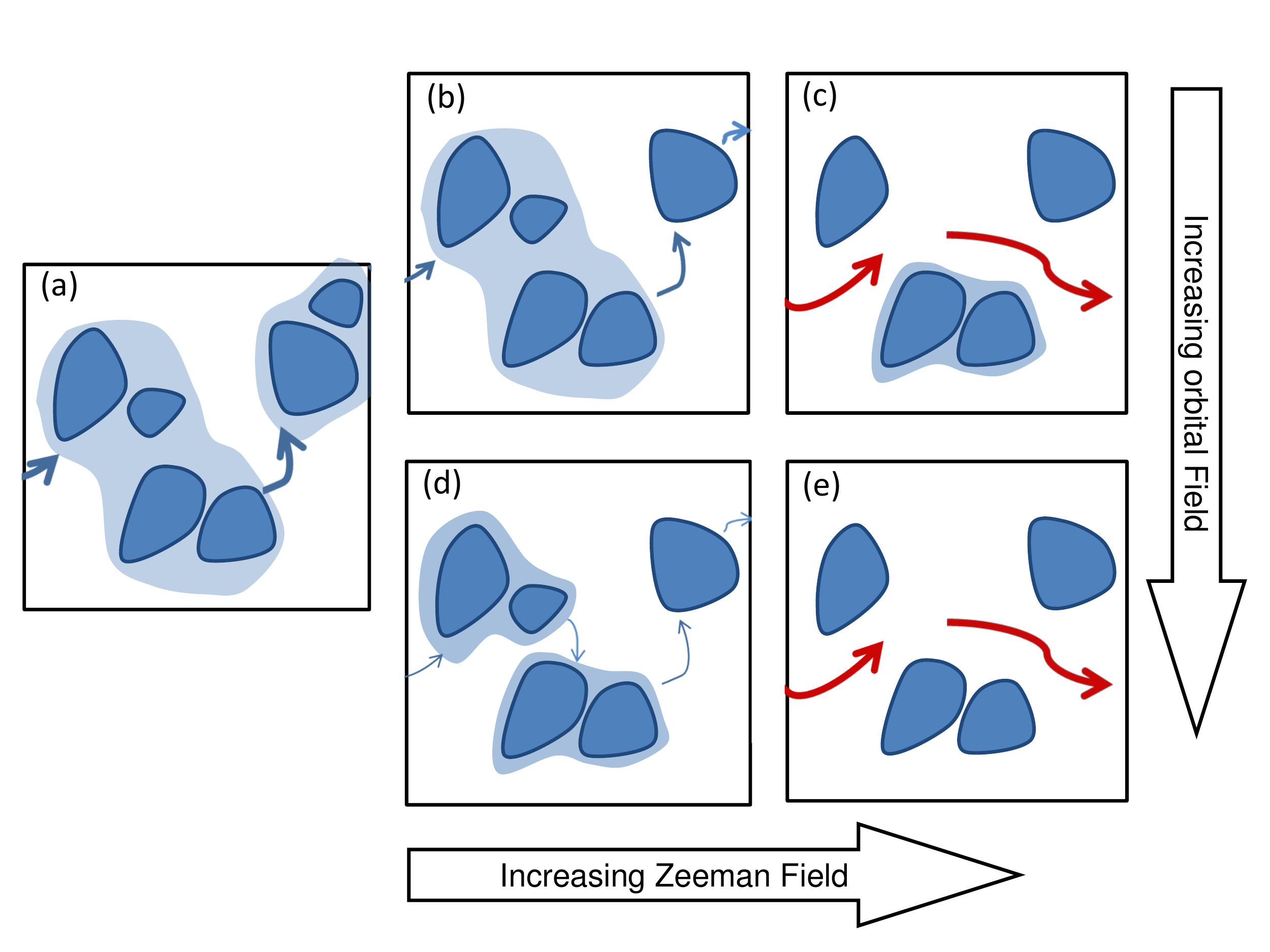}
\caption{\label{fig:schematic zemman model}Schematic representation of the
combined effects of the Zeeman and Orbital magnetic field: (a) At relatively low magnetic field, adjacent
SCIs (dark blue shapes) maintain phase coherence, thus forming
larger SC clusters (light blue aura). Most of the current is carried by SC
paths, i.e. paths which include SCIs (bold blue arrows). (b) As Zeeman
field is added, some of the smaller islands collapse and the total
resistance increases. (c) At sufficiently high Zeeman field the SC
paths become unfavorable compared to the purely normal paths (red
arrows), and normal current become dominant. From that point on isotropic or negative
MR is observed. (d) When orbital (perpendicular) field is added to
the Zeeman field of frame (b), some of the inter correlations between
adjacent SCIs are destroyed due to vortex penetration. This results
with increased total resistance, compared to parallel field of the
same amplitude. (e) When normal paths are dominant, the inter coherence
of adjacent SCIs, determined by the orbital field, becomes unimportant,
and the resistance becomes isotropic with respect to the direction
of the field.}
\end{figure}

The percolation theory of the MR of disordered SC thin films in perpendicular field is based on three assumptions \cite{Dubi2006}:
(1) The disorder induces fluctuations of the order parameter, which beyond the SIT results in formation of SCIs with nonzero pairing amplitude. This is supported by numerical \cite{Ghosal1998,Dubi2008} and experimental \cite{Kowal1994,Mason2001,Kowal2008,Sacepe2008,Kamlapure2013} data.
(2) Some of these SCIs are coherently coupled, forming a larger SC cluster. The concentration and size of these SC clusters are monotonically reduced under external magnetic field, presumably due to vortex penetration that destroys the coherence between the SCIs that form a single SC cluster. This results in a separation of the SC cluster into several smaller clusters. This picture is supported in numerical calculations \cite{Dubi2007}, taking into account phase correlations.
(3) Tunneling of electrons to the SCIs is suppressed due to a charging energy or  weak Andreev tunneling.

Under these assumptions, the SIT was interpreted \cite{Dubi2006} as a percolation transition, where the coherent SC clusters ceases to span the sample as their concentration is reduced below the percolation threshold. The MR peak was described as a crossover between Cooper-pair transport through the SCIs, via incoherent Josephson couplings \cite{Galitski2001}, to electron current, avoiding the SCIs because of the suppressed tunneling.
However, this theory, based on orbital effects of the magnetic field, cannot capture all of the recent observations. For example, if only orbital effects would matter, the MR curves at different angles would collapse onto each other upon appropriate scaling of the magnetic field, in contrast with observations (2)-(4). Here we include  two independent mechanisms,  orbital and Zeeman effects, whose relative importance, as will be shown below, varies with field amplitude and angle.
The interplay of these two field effects has already been investigated numerically in
\cite{Dubi2007,Dubi2008}, and the effect of the Zeeman field may be summarized by the following two conclusions:  (1) Increase of the Zeeman field results with consecutive and separate collapse of SCIs, due to the competition between the energy scales of the SC gap and the Zeeman energy \cite{Clogston1962,Chandrasekhar1962}. (2) Increase of orbital field decreases the average SC order parameter, thus allowing SCI collapse at smaller Zeeman field. It is the latter point, the fact that the impact of the Zeeman field depends also on the magnitude of the orbital field, which will lead to the reversed anisotropy at sufficiently high field.

The underlying physics is as follows. The disorder determines the concentration of the SCIs in the sample at zero field. These SCIs may be coherently coupled to form a large SC cluster, if the Josephson coupling is large enough to overcome quantum and temperature fluctuations. If a SC cluster percolates through the system, it is a superconductor, but with increasing orbital field, vortices penetrate the system, weakening the Josephson coupling between the SCIs, and eventually leading to loss of percolation, manifested by the SIT. This effect is highly anisotropic, due to the two-dimensionality of the system. SC order can also be lost by the isotropic Zeeman effect, that leads to the collapse of individual SCIs, when it exceeds the local SC gap.   At large fields, where the SCIs are few and small, the charge transport avoids the SCIs
due to the  tunneling cost. At that point the coherence among the SCIs,
determined by the orbital field, becomes irrelevant, and thus only the
isotropic Zeeman field, affecting the overall area of the SCIs, changes the resistance. This leads to the isotropic behavior seen at large fields.
A further increase of the orbital field would lower the typical SC gap, thus facilitating collapse of SCIs,  which  results with decreased resistance, and consequently with reversed anisotropy.
This physics is schematically demonstrated in Fig. \ref{fig:schematic zemman model},
where we show how an orbital field affects the resistance at low fields
(panels (b) and (d)), while having no effect at large fields ((c) and (e)).

\begin{figure}[hb]
\begin{centering}
\includegraphics[width=0.95\columnwidth]{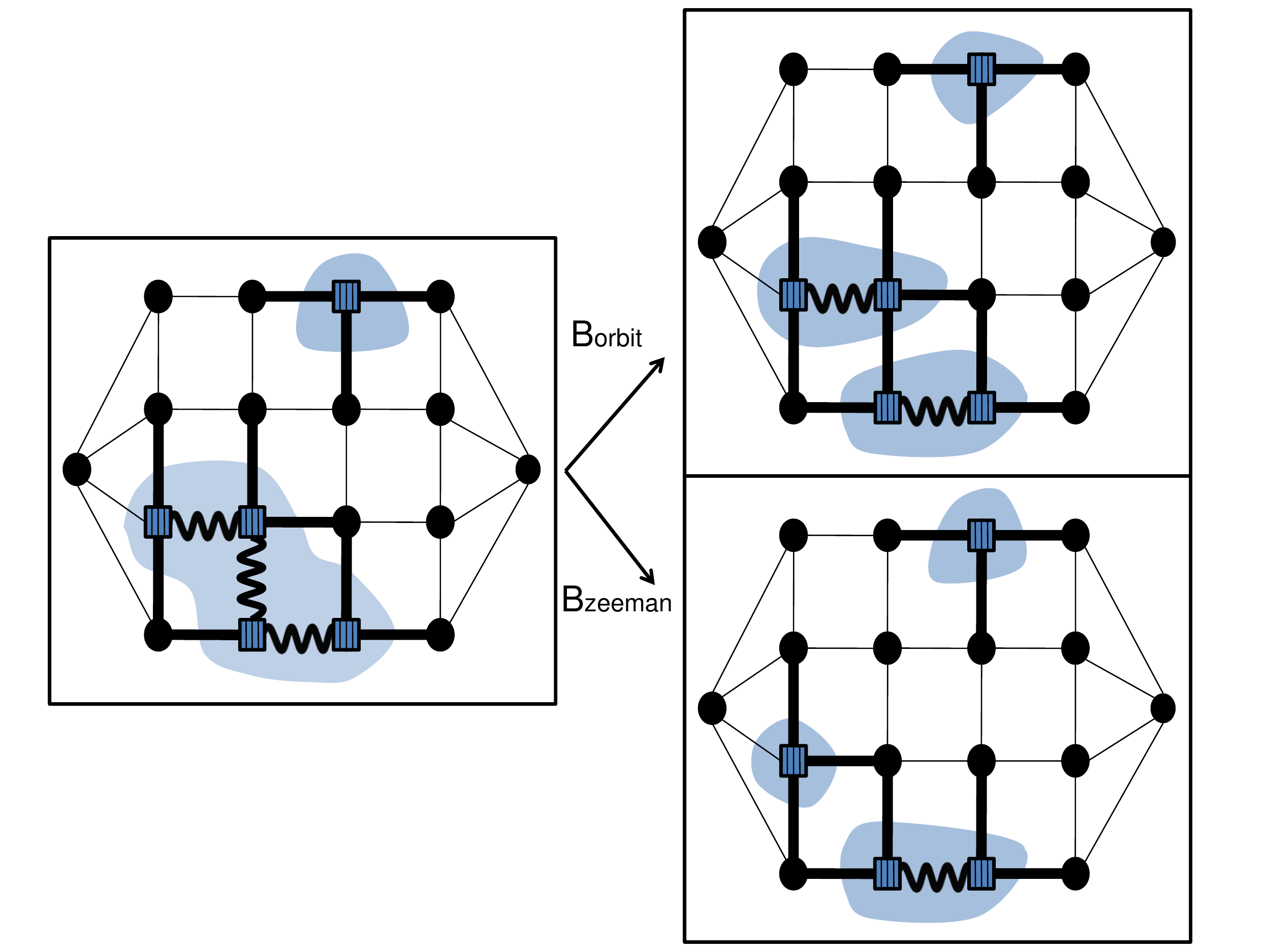}
\end{centering}
\caption{\label{fig:lattice-illustration}The site-bond lattice model. 
Left frame: The lattice sites represent either SCIs (squares) or normal regions
(small circles). The concentration of SC sites is determined by disorder
and the Zeeman field. Phase coherent SC sites are connected by SC
bonds (wavy lines), whose concentration is determined by the orbital
field. Correlated SC sites (connected by SC bonds) form SC islands (shaded areas).
A large blockade resistance (bold line) connects normal and SC sites.
As orbital field is added (top right frame), vortices penetrate the
sample and break the coherence between some adjacent SC sites. As
a result some of the SC bonds are broken and replaced with blockade
bonds, and SC clusters decompose into smaller clusters. On the other
hand, when Zeeman field is added (bottom right frame), some of the
SC sites collapse into normal sites, which results in an increase
of the normal paths.}
\end{figure}

To describe this physics quantitatively we introduce a site-bond percolation model, where the sites describe the SCIs and the links the coherence between them. We associate a uniform distribution of local gaps $P(\Delta_i)$ with the sites, such that any site with $\Delta_i>\Delta_0$ is considered SC (small squares in Fig.~\ref{fig:lattice-illustration}). Thus $\Delta_0$ describes the amount of disorder in the system. The Zeeman field
$B$ causes further destruction of SC sites, for which $\Delta_i-\Delta_{0}<B$. At zero field every nearest neighbor  SCIs are connected coherently forming a larger SC cluster (shaded areas in Fig. ~\ref{fig:lattice-illustration}). The orbital field $B_\perp\equiv B\sin\theta$, where $\theta$ is the angle between the field and the plane, has two effects: (a) Due to the penetration of vortices, the concentration of SC links ($p_{b}(B_\perp)$) decreases, causing a possible breakdown of a larger SC cluster into smaller ones, and (b) the orbital field affects $P(\Delta_i)$, the distribution of the local SC gap. An illustration of these effects of the orbital and Zeeman fields is presented in Fig. \ref{fig:lattice-illustration}. Quantitatively, to account for (a), we choose $p_b(B)=1-\Delta_0-B_{\perp}^{x}$. While the number of vortices is linear with the field, they will tend to penetrate the system and congregate in specific places, where the local gap is small. Thus the number of vortices that destroy SC links is effectively smaller, and we expect $x<1$. The same disorder parameter $\Delta_0$, that determines the concentration of the SCIs (sites in our lattice), also determines the concentration of the links connecting them at zero magnetic field.

The second effect of the orbital field is described in Ref. \onlinecite{Dubi2008} as a shift of the gap distribution towards zero. This is manifested in the model by a uniform shift of all the local gaps by the orbital field, $\Delta_i\rightarrow \Delta_i-\chi B_\perp^n$, where the susceptibility $\chi$ and the exponent $n$ are the parameters defining the dependence of the gap distribution on the orbital field.

In order to calculate the resistance of the sample, we assign resistances to the links. The resistance between nearest neighbor normal sites is activated \cite{Miller1960}, $R_{ij}=R_{0}\exp^{\left(\left|\epsilon_{i}\right|+\left|\epsilon_{j}\right|+\left|\epsilon_{i}-\epsilon_{j}\right|\right)/{k_{B}T}}$where $\epsilon_{i}$ is the energy of the site $i$, and $T$ is the temperature.
The site energies are taken from a uniform distribution $\left[-\nicefrac{W}{2},\nicefrac{W}{2}\right]$, 
where $W$ determines the disorder in energy.
On the other hand, the resistance of the SC links $R_{SC}(T)$ is taken to be very small compared to
$R_{0}$, and vanishes as $T\rightarrow0$. The precise functional
form of $R_{SC}(T)$ has no qualitative influence on the results,
and was arbitrarily taken to be a power law $R_{SC}\sim T^{\nicefrac{1}{2}}$. The resistance between SC and normal sites or between near neighbor uncorrelated SC sites  represents the charging energy required for electrons to enter the SCIs, thus their resistance is given by:
\begin{equation}
R_{B}=R_{B0}\exp\left[\tilde{E}_{c,i}/(k_{B}T)\right]\label{eq:Rb_T}
\end{equation}

where $\tilde{E}_{c,i}$ is the charging energy of the site,
and we used $R_{B0}=R_0$ throughout the calculations.
The charging energy  is expected to decrease with the SCI size. In the calculation we choose it to be  inversely proportional to the island size $\tilde{E}_{c,i}\sim E_{c}/S_{i}$ , in correspondence to charging energy of parallel plate capacitor, where
the cluster size $S_{i}$ is defined by the number of sites connected to site $i$ by SC bonds (i.e. cluster mass).
The calculation was also performed with other dependencies of $\tilde{E}_{c}$ on the size (including no dependence), and we find  that the MR is not qualitatively sensitive to the different
choices for the relation of $\tilde{E}_{c}$ and $S$.

Two representing results are displayed in Fig. \ref{fig:experiments}(c) and (d). The main features of the experimental data (Fig. \ref{fig:experiments}(a) and (b)) are clearly reproduced, including the strong anisotropy at low fields, that becomes weaker and is inverted at high fields, in accordance with the physics described above. The features are quite general, weakly dependent on model parameters, though the existence of a single
crossing point $B_{iso}$  in Fig. \ref{fig:experiments}(c) is only achieved for specific parameter choices.

\begin{figure}[t]
\includegraphics[bb=35bp 200bp 550bp 660bp,clip,width=0.98\columnwidth]{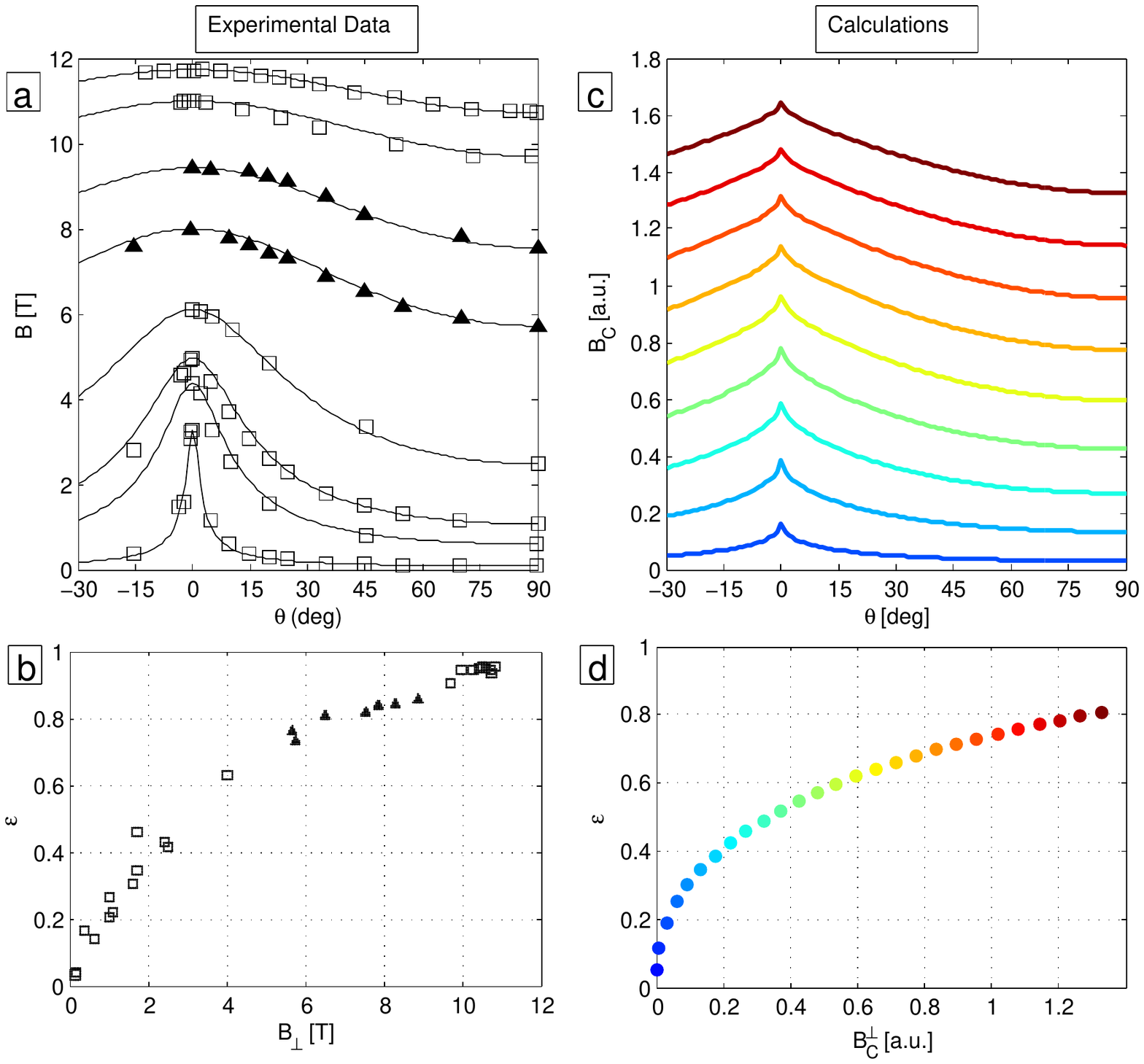}
\caption{\label{fig:anisotropy}
(a) Measured SIT critical field (empty squares) and peak field (filled triangles) of several films, for different orientations of B with respect to the plane of the film, taken from Ref. \onlinecite{Johansson2011}. 
The solid lines correspond to the theory of Ref. \onlinecite{Doniach1971}, which agrees with our theory with the parameter choice x=2.
(b) Isotropy factor $\epsilon\equiv\nicefrac{B_\perp}{B_{||}}$ as a function of $B_\perp$ of critical field
(empty squares) and peak field (filled triangles) of several films, taken from Ref. \onlinecite{Johansson2011}.
(c) Calculated orientation dependence of the critical field, as given by Eq. (\ref{eq:sitebond curve}) and our model assumptions, using the square 2D lattice pure site and bond thresholds: $P_C^s=0.593$,
$P_C^b=0.5$ and different values of disorder parameter $\Delta_0$, linearly varied between
-1 (dark red) to 0.4 (deep blue).
(d) Isotropy factor $\epsilon$ against $B_\perp$ as given by Eq. (\ref{eq:sitebond curve}), corresponding to angular data in (c). The points are colored with respect to the level of disorder, from $\Delta_0 = -1$ (dark red) to $\Delta_0 = 0.4$ (deep blue).}
\end{figure}

The experiment also determined the dependence of the critical magnetic field $B_c$ on tilt angle. In order to compare with these experimental observations, one does not need to employ the full resistor network calculation, but, in fact, 
we can utilize an approximate formula for the site-bond percolation critical curve \cite{Yanuka1990}:
\begin{equation}
\frac{\log p_s}{\log P^s_C}+\frac{\log p_b}{\log P^b_C}=1  \label{eq:sitebond curve}
\end{equation}
where $p_s, p_b$ are the critical site and bond concentrations for the site-bond percolation problem, and $P^{s(b)}_C$ are the pure site (bond) percolation thresholds. Substituting the dependence  of the site and bond concentrations on the magnetic field gives a transcendental equation for $B_c$:
$1-\Delta_0 - B_c = P^s_C\cdot(1-\Delta_0-(B_c\sin\theta)^x)^{-\alpha}$,
where $\alpha\equiv\nicefrac{\log P^s_C}{\log P^b_C}$. The latter equation can be solved numerically for the orientation dependence of the SIT critical magnetic field. In Fig. \ref{fig:anisotropy}(c), the angular dependence of the critical field is plotted for a different levels of disorder, represented by $\Delta_0$, and the critical thresholds used were those for the square 2D lattice. This result of the model is compared to the measured angular dependence of Ref. \onlinecite{Johansson2011} in Fig. \ref{fig:anisotropy}(a).
The qualitative similarity to the experiment is better emphasized in Fig. \ref{fig:anisotropy}(b) and \ref{fig:anisotropy}(d), where the isotropy factor, defined by $\epsilon\equiv\nicefrac{B^\perp_c}{B^{||}_c}$, is plotted against the parallel critical field, for the experimental and calculated data respectively. The calculations reproduce the convex, monotonic increase of isotropy with decreased disorder.

%
To conclude, we have addressed in this paper the magnetoresistance in thin superconducting disordered films as a function of the field direction. 
The experimental data provide both qualitative and quantitative constraints on possible
theories and may be crucial in pointing towards the correct theory that describes the superconductor-insulator transition and the huge magnetoresistance peak in the normal phase. In order to be able to explain the experimental features, and in particular the diminish of anisotropy with increasing field amplitude, one has to invoke both the orbital effect of the field and the Zeeman effect. Moreover, the interplay between the two was crucial to explain the high field behavior - the orbital field facilitates collapse of the superconducting islands by the Zeeman field, which at high fields suppresses the resistance. 
Rather unexpectedly, the magnetoresistance angle dependence suggests that the Zeeman effect has a non
negligible effect also at perpendicular fields of relatively small magnitude, particularly beyond the SIT.
The comparison between the experimental data and the numerical results raises a couple of interesting questions.
First, the zero-field insulating sample exhibits small negative MR at small parallel
field, which is currently not explained by our model. Second, while the the observation of a single isotropic
point can be explained by the numerical calculations for a limited set of parameters, it is not clear whether this is a coincidence, or that there are generic relations between the parameters of the model. Since such a crossing point has been reported in a single sample, we hope that this study will stimulate additional such experiments, in order to pinpoint the physics underlying the superconductor-insulator transition and the magnetoresistance peak.
We acknowledge discussions with M. Schechter, and support from the Israel Science Foundation (ISF).
\bibliographystyle{revtex}
\bibliography{paper_bib2}

\end{document}